\documentclass[12pt,dvips]{article}

\usepackage{epsfig}

\def\NPB#1#2#3{Nucl. Phys. {\bf B#1}, #2 (#3)}
\def\PLB#1#2#3{Phys. Lett. {\bf B#1}, #2 (#3)}

\def\PRD#1#2#3{Phys. Rev. {\bf D#1}, #2 (#3)}

\def\PRL#1#2#3{Phys. Rev. Lett. {\bf #1}, #2 (#3)}

\def\susy{{supersymmetry }}
\def\susc{{supersymmetric }}

\def\tb{{$\tan\beta$ }}

\def\GeV{{\,{\rm GeV}}}
\def\TeV{{\,{\rm TeV}}}
\def\u1{{U(1)}}
\def\u1x{{$\rm U(1)_F$}}
\def\su5{{SU(5)}}
\def\ldsb{{\Lambda_{D\!S\!B}}}

\def\gsm{{${\rm G}_{sm}$ }}
\def\gsmp{{${\rm G}_{sm}$}}
\def\mt{{\widetilde{m}^2}}
\def\huv{{\langle H^u \rangle}}
\def\hdv{{\langle H^d \rangle}}
\def\centeron#1#2{{\setbox0=\hbox{#1}\setbox1=\hbox{#2}\ifdim
        \wd1>\wd0\kern.5\wd1\kern-.5\wd0\fi
        \copy0\kern-.5\wd0\kern-.5\wd1\copy1\ifdim\wd0>\wd1
        \kern.5\wd0\kern-.5\wd1\fi}}
\def\ltap{\;\centeron{\raise.35ex\hbox{$<$}}{\lower.65ex\hbox{$\sim$}}\;}
\def\gtap{\;\centeron{\raise.35ex\hbox{$>$}}{\lower.65ex\hbox{$\sim$}}\;}

\def\lsim{\mathrel{\ltap}}

\begin{document}

\begin{titlepage}
\begin{flushright}
{UW/PT 98-07\\
DOE/ER/40561-12-INT98\\
CERN-TH/98-186}
\end{flushright}

\vskip 1.2cm

\begin{center}
{\Large\bf Fermion Masses and Gauge Mediated \\
        Supersymmetry Breaking from a Single U(1)}

\vskip 1.4cm

{\large
D. Elazzar Kaplan$^a$, Fran\c{c}ois Lepeintre$^{a,b}$, Antonio Masiero$^c$,\\
        Ann E. Nelson$^a$ and Antonio Riotto$^d$}
\vskip 0.4cm
{\it $^a$Department of Physics 1560, University of Washington,
                Seattle, WA 98195-1560\\
         $^b$Institute for Nuclear Theory 1550, University of Washington,
                 Seattle, WA 98195-1550\\
         $^c$SISSA, Via Beirut 2-4, 34013 Trieste, Italy, and INFN,
                 sez. Trieste, Italy\\
         $^d$Theory Division, CERN, CH-1211 Geneva 23, Switzerland
                \footnote{On leave  from Department of Theoretical Physics,
                        University of Oxford, U.K.}}

\vskip 4pt

\vskip 1.5cm

\begin{abstract}
We present a \susc model of flavor.
A single U(1) gauge group is responsible for both generating the
flavor spectrum and communicating supersymmetry breaking to the
visible sector.  The problem of Flavor Changing Neutral Currents is
overcome, in part  using an `Effective Supersymmetry' spectrum among the 
squarks, with the first two generations very heavy.  All masses are generated
dynamically and the theory is completely renormalizable.  The model contains a
simple Froggatt-Nielsen sector and communicates supersymmetry breaking via
gauge mediation without requiring a separate messenger sector.  By forcing the
theory to be consistent with \su5 Grand Unification, the model predicts a large
\tb and a massless up quark. While respecting the experimental bounds on CP
violation in the $K$-system,   the model leads to a large enhancement of CP
violation in $B-\bar B$ mixing as well as in $B$ decay amplitudes.
\end{abstract}

\end{center}

\vskip 1.0cm

\end{titlepage}

\section{Introduction}

Small dimensionless numbers in physics should have a known
dynamical origin \cite{thooft}.  However, Nature contains a number of
unexplained, seemingly fundamental small quantities, such as the ratio between
the weak scale and the Planck scale $\left(\frac{M_w}{M_p}\right)$, and the
ratios of known fermion masses to the weak scale
$\left(\frac{M_f}{M_w}\right)$.  The former is subject to large radiative
corrections in the Standard Model (SM).  But the hierarchy $M_w \ll M_p$
could be explained by dynamically broken supersymmetry with superpartner
masses near the weak scale, and the superpartner spectrum restricted such
that it satisfies experimental constraints on Flavor Changing Neutral
Currents (FCNC) and CP violation \cite{FCNC}.  By contrast, the fermion
masses are protected by an approximate chiral symmetry.  However, the SM
requires tiny dimensionless parameters to reproduce the measured spectrum.
These parameters could be produced dynamically by the spontaneous breaking
of a flavor symmetry.  A complete model would successfully predict the
entire spectrum of scalars and fermions with a Lagrangian that only
contained coupling constants of order unity.  In this article, we present
a model of supersymmetry and flavor which is renormalizable and
natural, and avoids excessive FCNC. All mass scales are generated
dynamically from the fundamental scale of supersymmetry breaking.

One way of mediating \susy breaking to the observable sector is
through gauge interactions \cite{GM82}.  In some of the first complete
models of gauge mediated \susy breaking (GMSB), a new gauge group,
$U(1)_{mess}$, couples to both a Dynamical Supersymmetry Breaking (DSB)
sector and a `messenger' sector to which \susy breaking is
communicated via loop effects \cite{DNS,DNNS}.  The messenger sector
consists of superfields that are vector-like with respect to the SM
gauge group (\gsm) and other superfields that are \gsm singlets.  At
least one \gsm singlet has a non-zero vacuum expectation value (vev)
with both scalar and auxiliary components, which in turn give \susc
and non-\susc masses respectively to the vector-like fields.  Squarks,
sleptons and gauginos receive \susy breaking masses from loop
corrections involving the messenger sector and SM gauge fields.  The
mass contributions come from gauge interactions and are therefore
flavor independent.  Hence, the three generations of scalars are very
nearly degenerate, naturally suppressing unwanted contributions to
FCNC.  Efforts to improve this scenario have been made in the last
few years, including attempts to remove the messenger sector and allow
the DSB sector to carry \gsm quantum numbers
\cite{GMrev}.

The most successful models of flavor are based on a mechanism developed by
Froggatt and Nielsen in the late 70's \cite{FN}.  In their original models, the
small Yukawa couplings of the SM are forbidden by an additional (gauged) \u1x
symmetry.  Quarks and leptons instead couple to Froggatt-Nielsen (FN) fields
(heavy fermions in vector-like representations of \gsmp), and scalar flavons,
$\phi$ (\gsm singlets).  Non-zero flavon vevs, $\langle\phi\rangle\ll M_{F}$
(where $M_{F}$ is the mass of the FN fields), break the \u1x and cause mixing
between the heavy and light fermions.  This produces Yukawa couplings in the
low energy effective theory proportional to the small ratio
$\epsilon\sim\frac{\langle\phi\rangle}{M_{F}}$ to some power $n$.  Here, $n$
depends on the charges of the relevant fermions.  A clever choice of charges
can produce the correct quark and lepton masses and quark mixing angles, all
with couplings of order unity.

These models of fermion masses and GMSB share a number of significant
features.  Both make use of an additional gauged U(1) symmetry which
is spontaneously broken, both contain heavy vector-like quarks and
leptons and both contain fields that are singlets under \gsmp.  These
similarities are striking and compel one to ask if these two
mechanisms can be incorporated efficiently into the same
model\footnote{These similarities were first noted by Arkani-Hamed,
  et al. \cite{arkani}.  In their article, they indicate some of the
  problems with identifying the two sectors.   These and other
  problems are addressed in this note.}.  There are, however, major
differences between the two mechanisms.  The biggest difference comes
from the fact that in the FN mechanism the vector-like fields and
some of the SM fields are charged under the U(1).  If the same were
true in GMSB, the squarks would not, in general, be degenerate.
However, large contributions to FCNC and CP violation can be
suppressed if the first two generations of squarks are very heavy, as
in ``Effective Supersymmetry''
\cite{dvali,CKN}.  If the first two generations carry U(1) charges,
their scalar components would be heavy due to loop effects, while
their fermion masses would be suppressed.  Models of this kind have
been built with the U(1) anomalies canceled at a high scale by the
Green-Schwartz mechanism
\cite{dvali,anomu1}.

In this article, we present a model that dynamically generates both
fermion and scalar masses using a single gauged U(1) which is non-anomalous.
In doing so, we employ a modified version of the FN
mechanism.  We produce the small ratio
$\epsilon\sim\frac{\langle\phi\rangle}{M_{F}}$ in a similar fashion.
However, the range of small parameters comes predominantly from the
use of flavons with different vevs producing different ratios as
opposed to different powers of the same ratio.  This method requires
fewer FN fields (at the cost of requiring more flavons), allowing us
to avoid a Landau pole in $\alpha_s$ below $M_{GUT}$.  While requiring
U(1) charge assignments to be consistent with \su5, we are able
to cancel all gauge anomalies, and
we are able to
find reasonable fermion mass matrices with fundamental
coupling constants of order unity.  The spectrum includes a massless
up quark, a viable solution to the strong CP problem.

The paper is laid out as follows:  Section 2 describes the overall
design of the model, the mass spectrum of the scalars and the
restrictions on the U(1) charges required for this spectrum.  Section
3 describes the fermion mass matrices allowed within these
restrictions.  Section 4 describes the contributions to FCNC and
shows that they fall within experimental bounds.  Section 5 
describes some interesting cosmological effects of the model, and 
Section 6 concludes the paper.  The Appendix shows why squarks cannot be
degenerate in this approach.

\section{Overview}

In this section, we describe the overall structure of the model.

\subsection{Supersymmetry Breaking} The highest scale defined in our model is
the one at which \susy breaks.  This breaking occurs in the DSB sector at
\begin{equation}
        \ldsb\sim 10^3-10^4\TeV .
\end{equation}
This scale is generated dynamically via nonperturbative effects.  Because there
are currently many types of models in which \susy is known or
believed to be broken dynamically \cite{ADS,dsb}, and because we have very few
requirements of this sector, we will leave it largely unspecified.  However,
the sector must contain a global U(1) symmetry which can be identified with a
$U(1)_{mess}$  gauge symmetry that communicates \susy breaking to the rest of
the model.  Once the DSB sector is integrated out, all lower scales will be
generated dynamically through radiative effects.

\subsection{Flavor and the Messengers of Supersymmetry Breaking}

\begin{figure}
\begin{center}
        \epsfig{file = 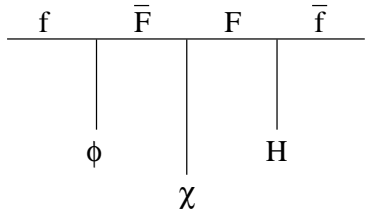}
        \caption{Source of $f$-${\bar f}$ mixing.}
        \label{fnmec}
\end{center}
\end{figure}

In order to naturally produce the small fermion masses of the SM, our model
contains Froggatt-Nielsen (FN) fields which are in vector-like representations
of \gsmp.  A \u1x gauge symmetry forbids most of the SM Yukawa couplings.  The
SM fields\footnote{When referring to `SM fields' we mean the superfields which
contain the standard model fields and their superpartners.} ($f,\bar{f}$)
instead couple to the FN fields ($F,\bar{F}$) and flavons ($\chi,\phi$) in the
superpotential
\begin{equation}
W\sim \chi F \bar{F} + \phi f \bar{F} + H f \bar{f}
\end{equation}
where $H$ is a Higgs superfield.  The scalar vev of $\chi$ produces a
mass term for the FN fields.  If $\phi$ has a scalar component with a
vev such that $\langle\phi\rangle\ll\langle\chi\rangle$, then the low
energy description of this theory will contain the superpotential term
$\sim\frac{\langle\phi\rangle}{\langle\chi\rangle}H f \bar{f}$ (see
Figure \ref{fnmec}).  Thus a small coupling is produced dynamically
from coupling constants of order unity.  Different small Yukawa couplings can
be produced by flavons with different vevs.  The \u1x charges are chosen so as
to produce fermion masses and mixing angles that mimic those experimentally
measured\footnote{Our model is `notationally' similar but significantly
different from another old and interesting approach to flavor by Dimopoulos
\cite{dim}}.

The DSB sector will also have fields charged under \u1x.
All other matter is assumed to couple to the DSB sector only via the
\u1x.
Fields carrying this charge will receive contributions to
their scalar masses at two loops.  By giving the first two generations
non-zero flavor charge, we can produce  the Effective Supersymmetry
spectrum \cite{CKN}.  The uncharged fields will be lighter and receive
their masses at one or two loops below $\ldsb$ (see
Section~\ref{scalars}).  The first two generations are heavy and adequately 
suppress unwanted contributions to FCNC and CP violation (Section~\ref{FCNC}).

\subsection{Flavor Symmetry Breaking}
We choose Froggatt-Nielsen fields that are vector-like under \gsm and chiral
under \u1x.  Their masses at tree-level will be proportional to flavon vevs
which break the flavor symmetry.  This symmetry breaking is due in part to a
Fayet-Iliopoulos (FI) term \cite{FI}, $\xi^2$, which appears in the \u1x
$D$-term:
\begin{equation}
\frac{g_{F}^2}{2}[\xi^2 + \sum_i q_i |\psi_i|^2]^2
\end{equation}
where $g_{F}$ is the gauge coupling and $q_i$ are the \u1x charges.
The fields, $\psi_i$ represent all charged fields, including both
trivial and non-trivial representations of \gsmp.   Provided that $\sum_i q_i$
vanishes, which is
necessary for anomaly cancelation, the FI term only receives finite
renormalization proportional to supersymmetry breaking effects.  We
assume that the fundamental FI term vanishes. Then the effective $\xi$ depends
on the DSB spectrum, and is generally an order of magnitude below $\ldsb$.

At two loops, every scalar with a non-zero $q_i$ receives a \susy
breaking mass squared proportional to its charge squared \cite{DNS,DNNS} 
\footnote{We assume there are no direct contact interactions between 
the DSB sector and the visible sector.}.  Specifically, the
contribution to the effective potential is $\mt\sum_i q_i^2 |\psi_i|^2$, where
the DSB sector again determines the exact value of $\mt$.  Its magnitude will
generally be two orders of magnitude below $\xi^2$.  Thus, after integrating
out the DSB sector, the full effective potential looks like
\begin{equation}
V_{e\!f\!f}=|\frac{\partial W}{\partial \psi_i}|^2 + \{ G_{sm}\: \mbox{D-terms}
\} + \frac{g_{F}^2}{2}[\xi^2 + \sum_i q_i |\psi_i|^2]^2 + \mt \sum_i q_i^2
|\psi_i|^2 + \cdots	\label{potential}
\end{equation}
where the ellipsis represent higher dimension \susy breaking terms.  
The \u1x D-term has a large number of flat directions.  The parameter
$\mt$ comes from the DSB and may have either sign.  As we will see in
Section~\ref{constaints}, the squared masses of the third generation and Higgs
scalars come from loop corrections which depend on $\mt$.  We find we must
have $\mt <0$ to keep squark masses positive.  This choice of sign introduces
runaway flat directions into Eq.~\ref{potential}.  These are curbed by the
higher dimension \susy breaking terms that we have ignored and by 
superpotential interactions.  We will choose a superpotential and a local
minimum that allows us to neglect the higher dimension terms.

How can we generate the appropriate flavon vev hierarchy?  One approach is to
give vevs only to $\chi$ fields at tree level. The $\phi$ flavons receive vevs 
at one or more loops.  Assume for instance that
the two flavons $\chi$ and $\chi'$ have vevs. The superpotential interaction
$\chi\chi'\phi$ gives a vev to the flavon $\phi$ via the diagram (solid and 
dashed lines represent fermion and scalar fields respectively):
\newline
\begin{center}
\includegraphics{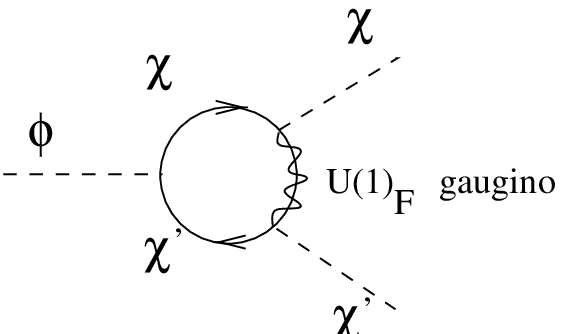}
\end{center}
Once $\phi$ has a vev, some other flavon $\phi'$ may receive a vev by means of
a similar diagram if it appears in the superpotential interaction
$\chi\phi\phi'$.  Such a technique produces a hierarchy of vevs.  In the above
case, for instance, $\left<\phi\right>$ and $\left<\phi^\prime\right>$ are
respectively one loop and two loop factors smaller than $\left<\chi\right>$.

Generating the hierarchy of vevs requires that we assign charges to the flavons
that allow the required superpotential interactions. It is also important to
prevent any field that transforms non-trivially under $SU(5)$ from acquiring a
vev.  Finally, additional flavons must be added to the model in order to cancel
the $U(1)_F$ and $U(1)_F^3$ anomalies.  Preliminary calculations have shown
that the above approach should yield a viable scalar potential.

\subsection{Mass Generation: Scalars}
\label{scalars}
As we have seen, all \u1x charged scalars have masses of at least order
$\tilde{m}$.  Uncharged scalars receive \susy breaking contributions from a
number of different sources.  Fields that transform non-trivially under \gsm
receive contributions from two loop diagrams in the low energy theory (below
$\ldsb$).  Drawing from the results of Poppitz and Trivedi \cite{PT97}, we find 
that the leading contribution to the mass  of an uncharged scalar at two loops
is (up to a group theory factor)
\begin{displaymath}
m^2_{unchg} \sim N \displaystyle \frac{\alpha_i^2}{2\pi^2} \Delta m^2 \log
\left(\frac{\Lambda^2_{DSB}}{m_f^2}\right)
\end{displaymath}
where $N$ is the number of charged Froggatt-Nielsen pairs, $i$ denotes the
relevant gauge group, $m_f$ is the fermionic mass of the Froggatt-Nielsen
fields and $\Delta m^2$ is of the order of the non-holomorphic contribution
to the scalar masses (i.e. $\Delta m^2 \sim \tilde{m}^2$).

The gaugino masses arise at one loop. Using again the results of \cite{PT97},
we find
\begin{displaymath}
m_{\tilde{g}} \simeq N \displaystyle \frac{\alpha_i}{4\pi} \frac{F}{m_f}
\end{displaymath}
where we have assumed that $F$ is significantly larger than $\Delta m^2$.
Here $\langle\chi\rangle = M + \theta\theta F$, where $\chi$ is a flavon
whose vev gives a mass to FN fields. Thus, $m_f=M$. These results assume
$F < M^2$, which is the case for our model. In order for the gauginos (and
in particular the winos) not to be too light compared to the lightest Higgs,
we require that $F$ be within an order of magnitude of $M^2$ (i.e.
$\frac{F}{M^2} > \frac{1}{10}$).  By choosing $\tilde{m}$ to be
about $20 \TeV$, we find that the light Higgs has a mass near the weak scale.

\begin{figure}
\begin{center}
        \epsfig{file=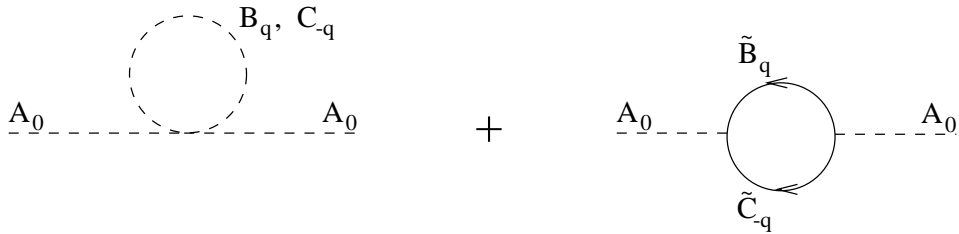}
        \caption{One loop contribution to the mass of an uncharged scalar, $A$,
                appearing in the superpotential term $W\sim A B C$.  The
                fields $B$ and $C$ have \u1x charges $q$ and $-q$
                respectively.}
        \label{loopunch}
\end{center}
\end{figure}

Uncharged fields with direct superpotential couplings to charged fields receive
scalar mass contributions from one-loop graphs containing charged fields
(Figure \ref{loopunch}) of order
\begin{equation}
\sim\frac{\lambda}{16\pi^2} {\tilde{m}^2} \log{\frac{\tilde{m}^2}{\ldsb^2}} ,
\end{equation}
where $\lambda$ is the superpotential coupling.  This contribution is
approximately an order of magnitude larger than the two-loop contribution
above.

The mass of an uncharged field may also receive a contribution from a charged
field due to $U(1)_F$ breaking if the charged and uncharged fields both appear
in the same F-term. For example, let us assume that the superpotential contains
$ABC + C\chi D$ where $A$, $B$ and $C$ are uncharged and $\chi$ and $D$ are
charged.  If $\chi$ has a non-zero vev, the squared masses of the scalar
components of $A$ and $B$ receive a contribution proportional to the
supersymmetry breaking mass of the scalar component of the charged $D$ field,
\begin{equation}
\sim\frac{\lambda}{16\pi^2}\tilde{m}^2.
\end{equation}
Moreover, if an uncharged field appears with a charged field in the same
F-term, they may mix due to \u1x breaking.  For example, the F-term
contribution to the scalar potential from the field $C$ above is $|A B +
D\chi|^2$.  If both $\chi$ and $B$ have non-zero vevs, then $A$ and $D$ would
mix.  The contributions described in this and the preceding paragraphs are not
flavor independent.  Thus, degenerate squarks are not a feasible method of
avoiding FCNC.

\subsection{Constraints on Charge Assignments and Couplings}
\label{constraints}
In choosing a Froggatt-Nielsen sector, our desire is to leave intact perhaps
the most compelling feature of the Minimal Supersymmetric Standard Model (MSSM), 
i.e., the unification of gauge
coupling constants.  To preserve this result, our vector-like FN fields should
come in complete \su5 representations.  In addition, \u1x charges should be
assigned to full multiplets.  Besides maintaining unification, this allows us
to satisfy easily the standard anomaly conditions as well as
\begin{equation}
{\rm Tr}[Y m_i^2]\simeq 0,
\end{equation}
where $m_i$ are scalar particle masses and Y is ordinary hypercharge.
If this equation were not satisfied, the $U(1)_Y$ $D$-term would
receive an unwanted Fayet-Iliopoulos term at one loop.

It is well-known that addition of
complete $SU(5)$ multiplets to the standard model does not ruin
coupling constant unification.
In order for the gauge couplings to remain perturbative from one-loop
running to the GUT scale, the following inequality must be satisfied:
\begin{equation}
3 n_{10} + n_{5} \stackrel{\textstyle <}{\sim} 5,
\end{equation}
where $n_{10}$ is the number of \{${\bf 10},{\bf\bar{10}}$\} pairs in
addition to the standard model fields, and $n_{5}$ is the number of
additional \{${\bf\bar{5}},{\bf 5}$\} pairs. Two loop contributions to
the beta functions will modify this condition, with two loop gauge
contributions
generally reducing slightly the number of additional fields allowed
and superpotential couplings increasing this number---we will assume
the net two-loop effects are not too important.
 A realistic model of fermion masses that satisfies this condition will have
$n_{10}=1$ and $n_{5}=1\mbox{ or }2$.  Thus, the particle content of our model
includes
\begin{itemize}
        \item{three generations of matter in \su5 multiplets, $\{ {\bf
10}_q^g,{\bf\bar{5}}_r^g \}$, where $g(=1,2,3)$ is the generation index, and
$q$ and $r$ denote \u1x charges,}
        \item{two Higgs superfields, $H^u$ and $H^d$,\footnote{The SU(5)
representations of the Higgs fields are intentionally left unspecified.  We do
not intend here to build a complete Grand Unified theory, but we wish to allow
unification to be possible in the context of our model.  We only require that
$H^u$ and $H^d$ contain the standard Higgs doublets.}}
        \item{Froggatt-Nielsen fields in vector representations of SU(5),
\newline$\{ {\bf 10}_d^V,{\bf{\bar 10}}_e^{\bar V}\}$, $\{ {\bf{\bar
5}}_l^V,{\bf 5}_m^{\bar V}\}$, and possibly $\{ {\bf{\bar 5}}_n^{V'},{\bf
5}_p^{\bar V'}\}$},
        \item{flavons (\su5 singlets) which have non-zero vevs -- some at
tree-level ($\chi$), and others at one or more loops ($\phi$), and}
        \item{additional fields $(A,B,C,\ldots)$ which help produce a 'cascade'
of flavon vevs}.
\end{itemize}

Another major constraint on the charge assignments of these fields
comes from the experimental limits on FCNC \cite{FCNC}.  There are
different ways to constrain squark (and slepton) masses in order to
limit \susc contributions to FCNC.  One way is to make their masses
degenerate, thus suppressing their contribution through a \susc GIM
mechanism.  Degeneracy is a natural result and thus a
virtue of the original GMSB models \cite{GM82,DNS,DNNS}.  In those models,
squark and slepton masses are dominated by loop corrections involving
flavor-blind \gsm couplings.  However, the additional structure in our
model produces significant flavor dependent contributions to sparticle 
masses, destroying this degeneracy.  Therefore, to suppress
FCNC, we instead decouple the problem by making the first two
generations heavy \cite{dvali,CKN}.  This can be achieved naturally by
simply requiring the particles in the first two generations, (${\bf
  10}_a^1,{\bf 10}_b^2,{\bf{\bar 5}}_i^1,{\bf{\bar 5}}_j^2$), to have
non-zero \u1x charges.  We do find, however, that some level of
degeneracy must still exist between the first two generations.

\begin{figure}
\begin{center}
        \epsfig{file=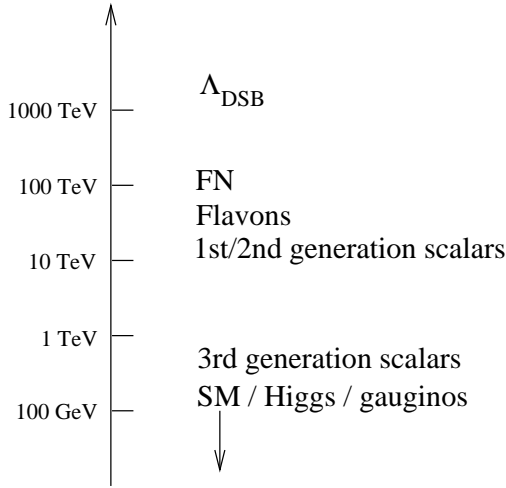}
        \caption{Spectral structure of the model.}
        \label{spectrum}
\end{center}
\end{figure}

The following observations impose additional constraints on our model:
\begin{itemize}
\item To avoid fine tuning, at least one Higgs must have a mass at the weak
scale.  Therefore, one Higgs must be uncharged (under \u1x) and must not have
any contact interactions with charged fields.
\item The higgsino mass will come from a $\mu$-type term in the superpotential,
\begin{equation}
W\supset X H^u H^d .
\end{equation}
Thus, to satisfy the previous condition, both Higgs fields must be uncharged.
\item The top quark's Yukawa coupling is of order unity and therefore does not
come from the Froggatt-Nielsen mechanism, but from a direct coupling to the
Higgs:
\begin{equation}
W\supset H^u {\bf 10}_c^3 {\bf 10}_c^3
\end{equation}
where c is the flavor charge and the 3 indicates the generation.
We conclude that $c=0$ by $U(1)_F$ invariance. Note also that $c=0$ guarantees 
that the $H^u$ mass contribution is not much larger than the weak scale.
\end{itemize}

Figure \ref{spectrum} summarizes the resulting spectrum.

\section{Fermion masses}
We want Yukawa coupling matrices in the low energy effective theory that
reproduce the known experimental values of fermion masses and mixing
angles.  In order to have a model from which the fermion masses of the SM
appear naturally, we must produce the small parameters in the Yukawa matrices
dynamically.  We accomplish this with a modified FN mechanism and a hierarchy
of flavon vevs.  This section describes the allowed fermion mass matrices.
\subsection{Framework}
The masses of the fermions are generated by superpotential terms like
$M_{ij}\psi_{i}\psi_{j}$, where $M_{ij}$ are the scalar vevs of Higgs or flavon
superfields.  To construct these superpotential terms, we apply the following
guidelines:
\begin{itemize}
\item We work in the context of \su5. This means our \u1x charge
  assignments are consistent with \su5.
\item We want the model to be natural. Any superpotential interaction should
appear with a coupling constant of order unity.
\item The Higgs fields are uncharged.  The up-type Higgs can not couple
directly to charged fields and the fields it couples to have restricted
interactions with charged fields.  The third generation ${\bf 10}$ is also
uncharged.
\item The first two generations must be charged in order to avoid large FCNC
(this will be shown explicitly).
\end{itemize}

{}From the following arguments, we will conclude that the FN sector must
include one \{${\bf 10}$, ${\bf\bar{10}}$\} pair and {\it two} \{${\bf 5}$,
${\bf\bar{5}}$\} pairs.  The model predicts a massless up quark and a large
value of $\tan\beta$.

The masses of up-type quarks come from the superpotential terms:
\begin{equation}
H^u {\bf 10\, 10}\;\;\mbox{ and }\;\;\varphi {\bf 10\, \bar{10}},
\end{equation}
while those of down-type quarks and leptons come from the terms
\begin{equation}
H^d {\bf \bar{5}\, 10},\;\:\varphi {\bf 10 \,\bar{10}}\;\mbox{ and }\;\varphi
{\bf \bar{5}\, 5}.
\end{equation}
Because of the SU(5) symmetry, the charged lepton mass matrix will be proportional
to the down quark mass matrix.  Deviations will derive from SU(5) breaking and
will depend on the Higgs sector of the (Grand Unified) model.  We will assume
that this can be done such that the correct lepton masses are predicted, and
thus for convenience, we shall speak only in terms of quark masses.

The quark content of the SU(5) multiplets are:
\begin{eqnarray*}
{\bf 10}_q^g &\supset& \{ \bar{u}_q^g , u_q^g , d_q^g \}\\
{\bf \bar{10}}_q^{\bar V} &\supset&
        \{ u_q^{\bar V},\bar{u}_q^{\bar V} , \bar{d}_q^{\bar V} \}\\
{\bf \bar{5}}_q^g &\supset& \{ \bar{d}_q^g \}\\
{\bf 5}_q^{\bar V} &\supset& \{ d_q^{\bar V} \},
\end{eqnarray*}
where $g(=1,2,3,V)$ is a generation index, and $q$ is the \u1x charge of the
multiplet.  Schematically, the tree-level mass matrices look like
\newline\newline
\begin{center}
\begin{tabular}{r|ccccc|}
&$\bar{u}_a^1$&$\bar{u}_b^2$&$\bar{u}_0^3$&$\bar{u}_d^V$&$\bar{u}_e^{\bar V}$\\
\hline
$u_a^1$ &\multicolumn{4}{c|}{ }&\\
$u_b^2$ &\multicolumn{4}{c|}{ }&$\langle$flav-\\
$u_0^3$ &\multicolumn{4}{c|}{$\langle\mbox{up-Higgs}\rangle$}&ons$\rangle$\\
$u_d^V$ &\multicolumn{4}{c|}{ }&\\ \hline
$u_e^{\bar V}${\rule[-3mm]{0mm}{8mm}}&\multicolumn{4}{c|}
        {$\langle\mbox{flavons}\rangle$}&$\hdv$\\
\hline
\end{tabular}
\end{center}
and
\newline
\begin{center}
\begin{tabular}{r|cccccc|}
&$\bar{d}_i^1$&$\bar{d}_j^2$&$\bar{d}_k^3$&$\bar{d}_l^V$&$(\bar{d}_n^{V'})$
&$\bar{d}_e^{\bar V}$\\
\hline
$d_a^1$ &\multicolumn{5}{c|}{ }&\\
$d_b^2$ &\multicolumn{5}{c|}{ }&$\langle$flav-\\
$d_0^3$ &\multicolumn{5}{c|}{$\langle\mbox{down-Higgs}\rangle$}&ons$\rangle$\\
$d_d^V$ &\multicolumn{5}{c|}{ }&\\  \hline
$d_m^{\bar V}${\rule[-3mm]{0mm}{8mm}}&\multicolumn{5}{c|}
        {$\langle\mbox{flavons}\rangle$}&$\huv$ \\
$(d_p^{\bar V'})$ &\multicolumn{5}{c|}{ }& \\
\hline
\end{tabular}
,
\end{center}
where the 6th row and 5th column of the down quark mass matrix represent the
optional (${\bf\bar{5}},{\bf 5}$) pair.  Now, following the above mentioned
guidelines on charge constraints, we can fill in these matrices.

Our strategy for avoiding large FCNC requires $a,b\neq 0$.  Therefore, any
field that appears in one of the first two rows of either matrix has a contact
interaction with a charged field.  However, the up-type Higgs field must not
interact with \u1x-charged particles, so the first two rows of the up matrix
will be devoid of Higgs vevs.  That matrix will have a zero eigenvalue, thus
predicting a massless up quark!  A vanishing up quark mass is a possible
solution to the strong CP problem, as the strong phase is no longer physical
and can be rotated into the up quark field via an axial rotation.  For complete
details on the viability of a massless up quark, see \cite{mu0}.

To complete the up matrix, we note that if $d\neq 0$, this matrix would have
two zero eigenvalues.  Since we are confident that the charm mass is not zero,
we set $d=0$.

Also, the Froggatt-Nielsen field $\bar{u}^V_0$ must interact with
$\bar{u}^{\bar{V}}_e$ through a flavon $\chi_{-e}$ to receive a mass
$\langle\chi_{-e}\rangle$ much greater than the weak scale. But $\bar{u}^V_0$
must interact with $H^u$ as well if the the up matrix has only one zero
eigenvalue. To avoid corrections to the up Higgs mass of order
$\displaystyle\frac{\tilde{m}}{4\pi}$ the fields interacting with $\bar{u}^V_e$
must be uncharged. That is, $e=0$.

Assuming that all allowed couplings exist, we find the up matrix is completely
determined and takes the form:
\begin{equation}
M^u=\bordermatrix{
        &\bar{u}_a^1&\bar{u}_b^2&\bar{u}_0^3&\bar{u}_0^V
                &\bar{u}_0^{\bar V}\cr
        u_a^1 &0&0&0&0&\langle \phi_{-a} \rangle\cr
        u_b^2 &0&0&0&0&\langle \phi_{-b} \rangle\cr
        u_0^3 &0&0&\huv&\huv&0\cr
        u_0^V &0&0&\huv&\huv&\langle \chi_{0} \rangle\cr
        u_0^{\bar V} &\langle \phi_{-a} \rangle&\langle \phi_{-b}\rangle
                &0&\langle \chi_{0} \rangle&\hdv\cr},
\end{equation}
where the fields ${\bf 10}_0^3$ and ${\bf 10}_0^V$ have been rotated to remove
the (3,5) and (5,3) entries.  For generic couplings, the (4,4) and (5,5)
entries have little effect on the final results.  For convenience, we
henceforth set them to zero\footnote{These couplings are relevant when dealing
with the '$\mu$-term problem.'  For details, see our Conclusion.}.  This matrix
produces the following up-type Yukawa couplings in the low energy theory:
\begin{eqnarray}
  \bar{u}
  \huv
    \left( \begin{array}{ccc} 0&0&\sim\!\epsilon_{1}\\ 0&0&\sim\!\epsilon_{2}\\
      \sim\!\epsilon_{1}&\sim\!\epsilon_{2}&\sim 1\\
      \end{array} \right)
  u,
\end{eqnarray}
where
\begin{displaymath}
\begin{array}{l}
\epsilon_{1}=\displaystyle\frac{\langle\phi_{-a}\rangle}{\langle\chi_{0}\rangle}\\
\epsilon_{2}=\displaystyle\frac{\langle\phi_{-b}\rangle}{\langle\chi_{0}\rangle}\\
\end{array}
\end{displaymath}
The tildes represent the (order 1) couplings that have not yet been included.

Now we shall attempt to design a down mass matrix with only one additional \{
${\bf\bar{5}}, {\bf 5}$\} pair.  First, to prevent a zero eigenvalue, there must 
be at least one $\hdv$ entry in one of the first two rows.  However, since we wish to
produce the small Yukawa couplings of the first two generations dynamically, we
place the entry in the 4th column.  To do this, we let $l=-b$ (choosing $-a$
would lead to the same conclusions).  Examining the first three columns, we see
that in order to avoid a zero eigenvalue, at least two of $i$, $j$ and $k$ must
be zero.  This is in contradiction with our decoupling strategy for avoiding
FCNC, hence ruling out this scenario.  One could ask if by setting all
$i=j=k=0$, these squarks would be degenerate.  However (see Appendix), the
degeneracy is broken by large flavor-dependent contributions.

We must include two ($\bar{5},5$) pairs in the FN sector.  Making similar
arguments as those above, we see our matrix is limited to
\begin{equation}
M^d=\bordermatrix{
        &\bar{d}_i^1&\bar{d}_j^2&\bar{d}_0^3&\bar{d}_l^V
                &\bar{d}_{-b}^{V'}&\bar{d}_0^{\bar V}\cr
        d_a^1 &0&0&0&?&0&\langle \phi_{-a} \rangle\cr
        d_b^2 &0&0&0&?&\hdv&\langle \phi_{-b} \rangle\cr
        d_0^3 &0&0&\hdv&?&0&0\cr
        d_0^V &0&0&\hdv&?&0&\langle \chi_{0} \rangle\cr
        d_m^{\bar V} &\langle \phi_{-i-m} \rangle
                &\langle \phi_{-j-m} \rangle&\langle \phi_{-m} \rangle
                &\langle \chi_{-l-m} \rangle&\langle \phi_{b-m} \rangle
                &0\cr
        d_p^{\bar V'} &\langle \phi_{-i-p} \rangle
                &\langle \phi_{-j-p} \rangle&\langle \phi_{-p} \rangle
                &\langle \phi_{-l-p} \rangle&\langle \chi_{b-p} \rangle
                &0\cr},
\end{equation}
where the question marks label undetermined entries.  We see that $l$ can be
either $(-a)$ or zero, and any of the flavons in the last two rows can be
removed.

\subsection{A Model}
\label{example}
We now present a specific example of the above framework that yields the
correct quark mass ratios and CKM angles.

If the up matrix is fixed, the down mass matrix would still allow many choices.
We start by choosing $l=-a$ ($l=0$ would work as well). The first four entries of
the fourth column of $M^d$ are then $\hdv$, 0, 0 and 0.  We want, for
simplicity, to limit the number of flavons appearing in the matrices. Again, we
can use the freedom offered by the down mass matrix. We can remove one flavon
from each of the first two columns -- the entries (6,1) and (5,2) are taken to
be 0.  We also can take the entries (5,5) and (6,4) to be 0.  As for the third
column one may ask if one could remove the two flavons in the entries (5,3) and
(6,3) since it would not generate a zero eigenvalue. However in such a scenario
the value of $V_{ub}$ comes out too small as is explained farther down.  We
take only entry (6,3) to be 0. The resultant matrices are
\\
\begin{equation}
\label{upmass}
M^{u} = \left(\begin{array}{*{5}{c}}
0&0&0&0&\langle\phi_{-a}\rangle \\
0&0&0&0&\langle\phi_{-b}\rangle \\
0&0&\huv& \lambda_1 \huv&0 \\
0&0&\lambda_1 \huv&0&\langle\chi_{0}\rangle \\
\langle\phi_{-a}\rangle&\langle\phi_{-b}\rangle&0&\langle\chi_{0}\rangle&0 \\
\end{array}\right)\end{equation}
\\
and
\\
\begin{equation}
\label{downmass}
M^{d} = \left(\begin{array}{*{6}{c}}
0&0&0&\mu_3 \hdv&0&\langle\phi_{-a}\rangle \\
0&0&0&0&\mu_2 \hdv&\langle\phi_{-b}\rangle \\
0&0&\hdv&0&0&0 \\
0&0&\mu_1 \hdv&0&0&\langle\chi_{0}\rangle \\
\langle\phi_{-m-i}\rangle&0&\langle\phi_{-m}\rangle&\langle\chi_{a-m}\rangle&0&0
\\
0&\langle\phi_{-p-j}\rangle&0&0&\langle\chi_{b-p}\rangle&0 \\
\end{array}\right),
\end{equation}
where the `$\lambda$'s and '$\mu$'s are coupling constants of order one that cannot
be absorbed by redefining the vevs.
Assuming $\langle\chi\rangle\gg\langle\phi\rangle$ and
$\langle\chi\rangle\gg\hdv,\huv$, we can integrate out the Froggatt-Nielsen fields, 
yielding the $3\times 3$ fermion mass matrices:
\begin{eqnarray*}
M^{u}_{3}=\huv  \left( \begin{array}{ccc} 0&0&\lambda_1 \epsilon_{1}\\
0&0&\lambda_1 \epsilon_{2}\\
     \lambda_1 \epsilon_{1}&\lambda_1 \epsilon_{2}&1\\
\end{array} \right)
\end{eqnarray*}
and
\begin{eqnarray*}
M^{d}_{3}=\hdv\left( \begin{array}{ccc} \mu_3 \epsilon_5&0&\mu_3 \epsilon_3 +
\mu_1 \epsilon_{1}\\
0&\mu_2\epsilon_4&\mu_1 \epsilon_2\\ 0&0&1\\
\end{array} \right)
\end{eqnarray*}
with
\begin{eqnarray*}
\epsilon_1=-\frac{\langle\phi_{-a}\rangle}{\langle\chi_{0}\rangle}&,&\epsilon_2=-\frac{\langle\phi_{-b}\rangle}{\langle\chi_{0}\rangle}\\
\epsilon_3=-\frac{\langle\phi_{-m}\rangle}{\langle\chi_{a-m}\rangle}&,&\epsilon_4=-\frac{\langle\phi_{-p-j}\rangle}{\langle\chi_{b-p}\rangle}\\
\epsilon_5=-\frac{\langle\phi_{-m-i}\rangle}{\langle\chi_{b-p}\rangle}&&
\end{eqnarray*}

We can now see why the (5,3) entry of $M^d$ cannot vanish.  The angle $V_{ub}$
is equal to the inner product $v_u^\dagger v_b$.  The vector $v_u$ is the
eigenvector of $M^{u}_{3}M^{u\dagger}_{3}$ corresponding to the eigenvalue
equal to the squared mass of the up quark (which is 0) and  $v_b$ is the
eigenvector of $M^{d}_{3}M^{d\dagger}_{3}$ corresponding to the eigenvalue
equal to the squared mass of the bottom quark. We have (up to some
normalization factors of order 1)
\begin{eqnarray*}
&&v_u=\left(\begin{array}{c}1\\-\displaystyle\frac{\epsilon_1}{\epsilon_2}\\0\end{array}\right)\\
&&v_b=\left(\begin{array}{c}\mu_1 \epsilon_1 + \mu_3 \epsilon_3 +
O(\epsilon^3)\\\mu_1 \epsilon_2 + O(\epsilon^3)\\\sim 1\end{array}\right)\\
\end{eqnarray*}
where $\epsilon$ is of the order of the $\epsilon_{i}$ in the matrices.
Typically, $\epsilon$ is less than $0.05$.
It follows:
\begin{displaymath}
V_{ub}=\mu_3 \epsilon_3 + O(\epsilon^3)\\
\end{displaymath}
If $\epsilon_3$ were 0, that is if there were no entry (5,3) in the $6
\times 6$ down matrix, $V_{ub}$ would be of order $10^{-4}$, an order
of magnitude too small to meet the experimental range. This short
computation also applies to the general form of the down mass
matrix. The ${\bf\bar{5}}^3$ field must always interact with one of
the ${\bf{5}}^V$ fields.  The mass of the right-handed (RH) bottom squark 
depends on this interaction. If ${\bf{5}}^V$ is charged, the mass of the 
RH bottom squark would be of order $\frac{\tilde{m}}{4\pi}$.  If 
${\bf{5}}^V$ is uncharged, the mass of the RH bottom squark would be at the weak
scale\footnote{The superpotential may contain couplings which contribute to a
right-handed bottom squark mass above the weak scale.}.

It remains to evaluate the orders of magnitude of the different
$\epsilon_{i}$s.  We find:
\begin{displaymath}
\left\{ \begin{array}{l}
\epsilon_2\simeq\sqrt{\displaystyle\frac{m_{charm}}{m_{top}}}\\
\\
\epsilon_1\simeq V_{us}\epsilon_2\\
\\
\epsilon_4\simeq\displaystyle\frac{m_{strange}}{m_{bottom}}\;\mbox{and}\;\epsilon_5\simeq\displaystyle\frac{m_{down}}{m_{strange}}\epsilon_4\\
\\
\epsilon_3\simeq V_{ub}\\
\\
\left(\mu_1 - \lambda_1 \right) \epsilon_2 \simeq V_{cb}\;\mbox{which
implies}\;\mu_1 - \lambda_1 \simeq \displaystyle\frac{1}{2}\\

\end{array}\right.\\
\end{displaymath}

\section{FCNC and CP violation}
\label{FCNC}

Several new interactions may contribute to $K^0-\bar{K}^0$ mixing and
$\epsilon_K$, beyond the usual  weak
interaction contributions. These usually provide the most stringent
constraints on supersymmetric models.  The  potentially largest
contribution
is from a gluino exchange box diagram:

\includegraphics{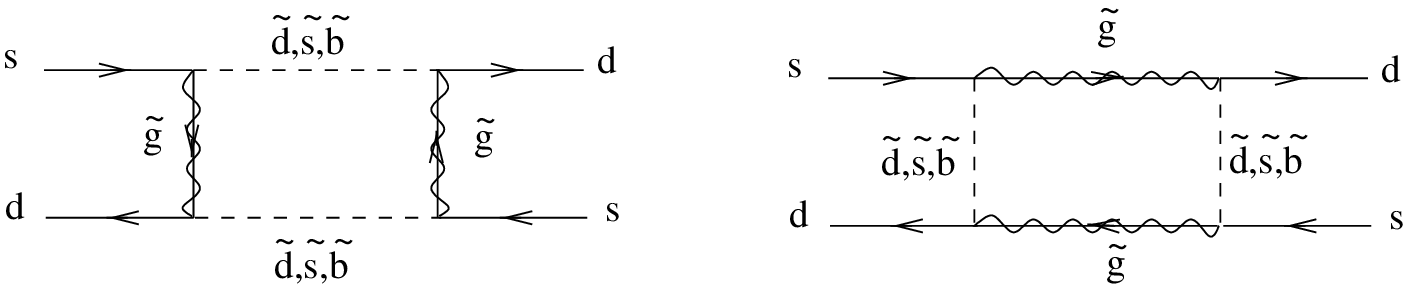}

We wish to compute in the framework of our model the two main contributions to
$K^0-\bar{K}^0$, namely the contribution of the quarks via the usual weak
interactions and the contribution from the squarks due to the strong
interactions.  We work with the specific model described in
Section~\ref{example}. In this example,
we shall find that the squark contribution to
the $K_S-K_L$ mass difference is
small,  both because the first two squark generations are heavy, and
because the squark mixing  angles amongst the first two generations of
the down sector are very small \cite{NS}. The
largest supersymmetric contribution  comes from the left handed bottom
squark. The phase of this contribution is naturally almost real, and
so the contribution to $\epsilon_K$ is sufficiently small.

To generate CP violation, there must exist at least one complex parameter in
the interaction Lagrangian that cannot be made real by redefining fields. The
relevant interactions are listed in the up and down mass
matrices~(\ref{upmass}) and~(\ref{downmass}). Assuming that all coupling
constants and vevs are complex, suitable redefinitions of the fields that do
not receive vevs allow us to make some of the entries in the mass matrices
real. However, two entries cannot be made real. We can choose these to be the
(3,3) and (5,3) entries of the down matrix.  The up and down matrices are then
(ignoring real coupling constants of order 1)
\\
\begin{displaymath}
M^{up} = \left(\begin{array}{*{5}{c}}
0&0&0&0&\langle\phi_{-a}\rangle \\
0&0&0&0&\langle\phi_{-b}\rangle \\
0&0&\huv& \huv&0 \\
0&0& \huv&0&\langle\chi_{0}\rangle \\
\langle\phi_{-a}\rangle&\langle\phi_{-b}\rangle&0&\langle\chi_{0}\rangle&0 \\
\end{array}\right)\end{displaymath}
\\
and
\\
\begin{displaymath}
M^{down} = \left(\begin{array}{*{6}{c}}
0&0&0&\hdv&0&\langle\phi_{-a}\rangle \\
0&0&0&0&\hdv&\langle\phi_{-b}\rangle \\
0&0&\eta_{33}\hdv&0&0&0 \\
0&0&\hdv&0&0&\langle\chi_{0}\rangle \\
\langle\phi_{-m-i}\rangle&0&\eta_{53}
\langle\phi_{-m}\rangle&\langle\chi_{a-m}\rangle&0&0 \\
0&\langle\phi_{-p-j}\rangle&0&0&\langle\chi_{b-p}\rangle&0 \\
\end{array}\right),
\end{displaymath}
where $\eta_{33}$ and $\eta_{53}$ are two complex numbers of modulus 1.

All of the other variables in the matrices are real.  The corresponding
superpotential reads
\begin{eqnarray}
\label{potexa}
W_{FN}&=&H^d 10_a^1 \bar{5}_{-a}^V + \phi_{-a} 10_a^1 \bar{10}_0^{\bar{V}} +
\phi_{-b}10^2_b\bar{10}_0^{\bar{V}} \nonumber \\
&+&H^d 10^2_b \bar{5}_{-b}^{V^\prime} + \eta_{33}H^d 10_0^3 \bar{5}_0^3  + H^d
10_0^V \bar{5}_0^3 \nonumber \\
&+&\chi_0 \bar{10}_0^{\bar{V}} 10_0^V + \phi_{-m-i} \bar{5}_i^1 5_m^{\bar{V}} +
\eta_{53}\phi_{-m} \bar{5}_0^3 5_m^{\bar{V}}\\
&+&\chi_{a-m} 5_m^{\bar{V}} \bar{5}_{-a}^V + \phi_{-p-j} \bar{5}_j^2
5_p^{\bar{V}^\prime} + \chi_{b-p} 5_p^{\bar{V}^\prime} \bar{5}_{-b}^{V^\prime}
\nonumber\\
&+&\frac{1}{2} H^u 10_0^3 10_0^3 + H^u 10_0^3 10_0^V \nonumber \\
\nonumber \end{eqnarray}
Integrating out the heavy fields, we find the following up and down matrices
for the light fermions:
\begin{eqnarray*}
M^{u}_{3}=\huv  \left( \begin{array}{ccc} 0&0&\epsilon_{1}\\ 0&0&\epsilon_{2}\\
     \epsilon_{1}& \epsilon_{2}&1\\
\end{array} \right)
\end{eqnarray*}
and
\begin{eqnarray*}
M^{d}_{3}=\hdv\left( \begin{array}{ccc} \epsilon_5&0&\eta_{53}\epsilon_3 +
\epsilon_{1}\\
0&\epsilon_4& \epsilon_2\\ 0&0&\eta_{33}\\
\end{array} \right)
\end{eqnarray*}
from which we get the CKM matrix:
\begin{eqnarray*}
V^{CKM} = \left(\begin{array}{ccc}
1&-\displaystyle\frac{\epsilon_{1}}{\epsilon_{2}}&-\eta_{33}^*\eta_{53}\epsilon_{3}\\\displaystyle\frac{\epsilon_{1}}{\epsilon_{2}}
&1&\left(\eta_{33}^*-1\right)\epsilon_{2}\\
\left(1-\eta_{33}\right)\epsilon_{1}&\left(1
-\eta_{33}\right)\epsilon_{2}&1\\ \end{array}\right).
\end{eqnarray*}
Only the significant phases have been retained.  The phases of
$\eta_{33}$ and $\eta_{53}$ are assumed to be of order 1.  The
remaining entries have phases of order $10^{-2}$ or less.  Such a CKM
matrix yields reasonable values of $\Delta m_{K}$ and $\epsilon_K$
from the weak interactions.

The contribution of the gluino box to $K^0-\bar{K}^0$ mixing remains to be computed.
To compute this requires the squark mass matrix. We consider tree level and one
loop mass terms generated by the effective scalar potential.

We assume that all of the flavons appearing in one line or column of the mass
matrices are distinct (this is automatically satisfied if all the standard
model fields have different charges). This implies that there are no off-diagonal 
one-loop corrections to the squark mass matrix of order:
\begin{displaymath}
\displaystyle\frac{\tilde{m}^2}{16\pi^2}\log\left(\frac{\Lambda_{DSB}}
        {\langle\chi\rangle}\right)
\end{displaymath}
Indeed, if for example $a=b$, we could have $\phi_{-a}=\phi_{-b}$. The F-term
of $\bar{10}^{\bar{V}}_0$ would yield the interaction:
\begin{displaymath}
{\bf 10}^1_{a} {\bf 10}^{2\,\ast}_{a} \phi_{-a} \phi_{-a}^{\ast}
\end{displaymath}
from which we could get the one-loop scalar graph:
\begin{center}
\includegraphics{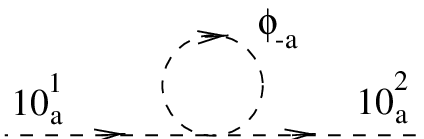}
\end{center}
whose supersymmetry breaking part is of the order of the above correction.
The only one loop corrections to off diagonal terms come from the supersymmetry
breaking part of scalar graphs like:
\begin{center}
\includegraphics{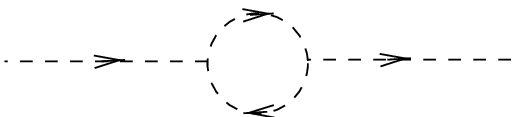}
\end{center}
Both left-handed squarks (LH) and right-handed squarks (RH) will contribute to
these processes.

We first consider the case of the LH down squarks, since as we shall see their
contribution is the largest. At tree level, the masses of the LH down
type squarks come from the hermitian matrix (omitting the vev symbol
$\langle\rangle$ for clarity):
\begin{displaymath}
\bordermatrix{
&d_a^{1\,\ast} &d_b^{2\,\ast} &d_0^{3\,\ast} &d_0^{V\,\ast}
        &d_m^{\bar{V}\,\ast}&d_p^{\bar{V}^\prime\,\ast}\cr
d_a^1& \tilde{m}^2& \phi_{-a}\phi_{-b}& 0 &\phi_{-a}\chi_0&\chi_{a-m} H^d&0\cr
d_b^2&    & \tilde{m}^2& 0&\phi_{-b}\chi_0&0&\chi_{b-p}H^d\cr
d_0^3&   &   &m_\mathit{weak}^2&\eta_{33} H^{d\,2}
        &\eta_{33}\eta_{53}^{\ast}\phi_{-m} H^d&0\cr
d_0^V& & & &\chi_0^2&\eta_{53}^{\ast} \phi_{-m} H^d&0\cr
d_m^{\bar{V}}& & & & &\chi_{a-m}^2&0\cr
d_p^{\bar{V}^\prime}& & & & & &\chi_{b-p}^2\cr}
\end{displaymath}
where we have written only the main contribution to each matrix element.
\newline
There are other tree level contributions to the masses of the LH down squarks
since some mixing occurs with the RH down squarks.  However, as we note at the
end of the present section, these terms are small and can be ignored for an
order of magnitude computation.

As mentioned above, any off-diagonal entry of the mass matrix may receive a 
one-loop contribution. The correction is of order
\begin{displaymath}
\displaystyle\frac{1}{16\pi^2}\langle A\rangle\langle
B\rangle\log\left(\displaystyle\frac{m_{\mathit{fermion}}^2}{m_{\mathit{scalar}}^2}\right)
\end{displaymath}
The masses appearing in the $\log$ are the masses of the heavier scalar
particle running in the loop and of its fermionic partner.

Some loop corrections come from known superpotential interactions between
$SU(5)$ and flavon fields in~(\ref{potexa}).  For instance, from the F-terms of
$\bar{5}_0^3$ and $\bar{5}_{-a}^V$, we get a diagram that mixes $d_a^1$ with
$d_0^3$:
\begin{center}
\includegraphics{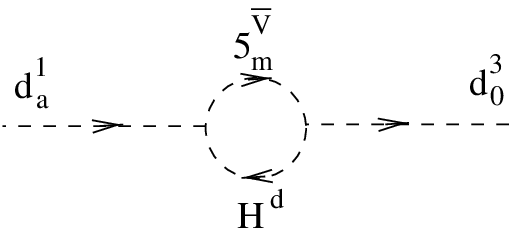}
\end{center}
The resulting term is
\begin{displaymath}
\displaystyle\eta_{53}\eta_{33}^{\ast}\frac{1}{16\pi^2}\langle\phi_{-m}\rangle\langle\chi_{a-m}\rangle\frac{\tilde{m}^2}{\langle\chi_{a-m}\rangle^2}
\end{displaymath}

Other loop corrections may arise from terms in the flavon superpotential.
Without knowing explicitly the flavon superpotential, we cannot tell if one
specific entry receives a correction and if so what the vevs $\langle A\rangle$
and $\langle B\rangle$ are. We will assume that any off-diagonal term in the
matrix receives such a correction with a phase of order 1 and that the vev
product is of the order of $\langle\phi\rangle\langle\chi\rangle$ with
$\displaystyle\frac{\langle\phi\rangle}{\langle\chi\rangle}\simeq 10^{-2}$.
This last value is an overestimate (most likely the vev product is of the order
 of the product of two $\phi$ vevs).  For example, assuming a flavon
superpotential containing the terms $CD\phi_{-a}$ and $CD^\prime\phi_{-b}$, we
obtain an off-diagonal term mixing $d_a^1$ and $d_b^2$. The loop correction is
equal to $\displaystyle\sim\frac{1}{16\pi^2}\langle D\rangle^\ast\langle
D^\prime\rangle\frac{\tilde{m}^2}{\langle\chi_{a-m}\rangle^2}$. We assume that
${\langle D\rangle}^\ast \langle D^\prime\rangle$ is of the same order as
$\langle\phi\rangle\langle\chi\rangle$.

We may now estimate the angles at the squark-quark-gluino vertex with the
quarks and squarks taken as mass eigenstates. For the LH quarks, the angles are
given by the matrix that diagonalizes $M^{d}_{3}M^{d\,\dagger}_{3}$:
\begin{eqnarray*}
\left( \begin{array}{ccc} 1& -(\eta_{53}^{\ast}\epsilon_3 +
\epsilon_1)\epsilon_2&\eta_{33}\epsilon_1\\
(\eta_{53}\epsilon_3 + \epsilon_1)\epsilon_2&1&-\eta_{33} \epsilon_2\\
-\eta_{33}^{\ast}\epsilon_1&\eta_{33}^{\ast}\epsilon_2&1\\
\end{array} \right) \sim
\left( \begin{array}{ccc} 1&10^{-3}&10^{-2}\\
                           & 1  & 10^{-1}\\
                           & & 1\\
\end{array} \right)
\end{eqnarray*}
Off-diagonal elements in the third line and column have the same order one
phase.  The elements (1,2) and (2,1) have a smaller phase of order $10^{-1}$.

For the squarks, we first integrate out the heavy fields and then rotate the
light squark mass matrix. Doing so, we find the following symmetric matrix as 
an estimate for the rotation matrix:
\begin{eqnarray*}
\left( \begin{array}{ccc} 1&\epsilon_1 \epsilon_2
\displaystyle\frac{\langle\chi\rangle^2}{\tilde{m}^2}
&\displaystyle\frac{1}{16\pi^2}\frac{\langle\phi\rangle}{\langle\chi\rangle}\\
 & 1 &
\displaystyle\frac{1}{16\pi^2}\frac{\langle\phi\rangle}{\langle\chi\rangle}\\
 & & 1\\
\end{array} \right) \sim
\left(\begin{array}{ccc}1&10^{-2}&10^{-4}\\
                        &1&10^{-4}\\
                        &&1\\
      \end{array}\right)
\end{eqnarray*}
where the off-diagonal elements of the third line and column have phases  of
order 1 and the entries (1,2) and (2,1) have a phase of order $10^{-2}$.  The
vev $\langle\chi\rangle$ is a generic value for the vevs of $\chi_0$,
$\chi_{a-m}$ and $\chi_{b-p}$, which we assume to be all of the same order. The
value of $\displaystyle\frac{\langle\chi\rangle}{\tilde{m}}$ depends on the
DSB.  A typical value is
$\displaystyle\frac{\langle\chi\rangle^2}{\tilde{m}^2}=10$.  As before,
$\displaystyle\frac{\langle\phi\rangle}{\langle\chi\rangle}$ is taken to be
$10^{-2}$.

The product of the above matrices yields the matrix $Z_{LL}$ at the
squark-quark-gluino vertices
\begin{displaymath}
Z_{LL} \sim \left( \begin{array}{ccc} 1&10^{-2}&10^{-2}\\
                             &1&10^{-1}\\
                             &&1\\ \end{array} \right),
\end{displaymath}
where off-diagonal elements (1,3) and (2,3) have phases which are of order 1 but
differ by a term of order $10^{-2}$. The (1,2) angle has a phase of $10^{-2}$.

The contribution of the LH squarks alone to the box diagram is \cite{FCNC}:
\begin{displaymath}
\displaystyle\langle \bar{K} | H_{LL} | K \rangle = \frac{1}{3}\alpha_s^2
Z^{1i\,\ast}_{LL} Z^{2i}_{LL} Z^{1j\,\ast}_{LL}
Z^{2j}_{LL}\left(-\displaystyle\frac{11}{36}I_1 +
\displaystyle\frac{x_g}{9}I_2\right)\displaystyle\frac{1}{\tilde{m}^2} m_K
 f_K^2,
\end{displaymath}
where
\begin{eqnarray*}
I_1 = \int_0^{+\infty}dy\,
\displaystyle\frac{y^2}{\left(y+x_i\right)\left(y+x_j\right)\left(y+x_g\right)^2}&&\\
I_2 = \int_0^{+\infty}dy\,
\displaystyle\frac{y}{\left(y+x_i\right)\left(y+x_j\right)\left(y+x_g\right)^2}&&\\
\end{eqnarray*}
and
\begin{eqnarray*}
x_i = \frac{m_i^2}{\tilde{m}^2}&&\\
\end{eqnarray*}
The indices i and j refer to the squarks and run from 1 to 3. The index g
stands for the gluino, whose mass is at the weak scale ($x_g\simeq 10^{-4}$).
Parameters $m_K$, $f_K$ and $\alpha_s$ are the K mass, K decay constant and 
strong coupling constant ($m_K=490\:\mathit{MeV}$, $f_K=160\:\mathit{MeV}$ and
$\alpha_s\left(M_W\right)=0.12$).

Given the above $Z_{LL}$ matrix and taking $\tilde{m}\simeq 20\TeV$, we find all
contributions to $\Delta m_K$ and $\epsilon_K$ are within the experimental
values. For instance, a LH bottom squark of mass $m_{\mathit{weak}}$ gives a
contribution to $\Delta m_K$ of $10^{-13}\,\mathit{MeV}$ and to $\epsilon_K$ of
$10^{-3}$. Other possibilities involving first or second generations squarks
give smaller contributions.

The same computation done with the RH down quarks and squarks give the
following matrix $Z_{RR}$ (again neglecting left-right mixing):
\begin{displaymath}
Z_{RR} \sim \left( \begin{array}{ccc} 1&10^{-4}&10^{-4}\\
                             &1&10^{-3}\\
                             &&1\\ \end{array} \right),
\end{displaymath}
where all off diagonal entries have a phase of order 1. This matrix gives
contributions to $\Delta m_K$ and $\epsilon_K$ well below the experimental
bounds.

We now consider the mixing between LH and RH squarks and confirm
that it can be neglected.

At tree level, a mu term $\mu H^u H^d$ generates mixing between $d^3_{0}$ and
$\bar{d}^3_{0}$ and also between the light squarks, $\bar{d}^3_{0}$, $d^1_{a}$,
and $d^2_{b}$, and heavy squarks, namely $d_{0}^{V}$, $\bar{d}^V_{-a}$ and
$\bar{d}^{V^\prime}_{-b}$. These terms come with a coefficient $\mu \huv \simeq
m_{\mathit{weak}}^2$ and possibly a phase of order 1. They are of the same
order as the loop corrections previously considered and would not change the
order of magnitude of $Z_{LL}$ and $Z_{RR}$.

Additional mixing may occur due to the flavon superpotential. For instance, the
flavon $\phi_{-b}$ could appear in the flavon superpotential in a term
$\phi_{-b} D D^{\prime}$. Assuming that $D$ and $D^{\prime}$ receive a vev, a
mixing term between $d^2_{b}$ and $\bar{d}_{0}^{\bar{V}}$ is generated. Its
coefficient is $c_{b}=\langle D\rangle^{\ast}\langle D^{\prime}\rangle^{\ast}$.
Similarly, via $\phi_{-a}$, we discover a mixing term between $d^1_{a}$ and
$\bar{d}_{0}^{\bar{V}}$ could also be generated (with coefficient $c_a$ equal
to a product of flavon vevs). This would mean that when integrating out
$\bar{d}_{0}^{\bar{V}}$, the entry (1,2) of $Z_{RR}$ would receive a
 contribution $\frac{c_a c_b}{\langle\chi\rangle^2\tilde{m}^2}$.
 The previous analysis could be invalidated if $c_a$ and $c_b$ were large, 
causing contributions to FCNC and CP violation beyond experimental bounds.
We must constrain the choice of the flavon superpotential. We assume that the 
vev product $\langle D\rangle\langle D^{\prime}\rangle$ is of the order of
$\langle\chi\rangle\langle\phi\rangle$ with a small phase ($\simeq 10^{-2}$) or
of the order of $\langle\phi\rangle^2$ with a phase of order 1. If so, the
orders of magnitude of $Z_{LL}$ and $Z_{RR}$ are unchanged. These restrictions
are reasonable since most $\phi$ fields do not interact directly with a $\chi$
field.

With these restrictions, left-right angles remain small (of the order of
$10^{-3}$ or less) except for the mixing angle between $d^3_{0}$ and
$\bar{d}^3_{0}$ which is about $10^{-2}$. The contribution to the gluino box of
left-right effects is then well below the experimental values.

We may conclude that the framework of our model accommodates the current
experimental bounds on FCNC and CP violation for the $K$ system. The important
point in this analysis was to assume that there were no loop corrections to 
off-diagonal entries given by
\begin{displaymath}
\displaystyle\frac{\tilde{m}^2}{16\pi^2}\log\left(\frac{\Lambda_{DSB}^2}{{\langle\chi\rangle}^2}\right)
\end{displaymath}
which would generate angles of order $10^{-1}$. With such a correction, FCNC
are still within the experimental values. However, CP violation would be much
larger than the experimental bound if the correction came with a phase of order
1.

This analysis is applicable to the $B$-system.  The entries in the
third column of $Z_{LL}$ are large and with a phase of order 1 and the
mass of the left-handed bottom squark is at the weak scale.
Therefore, the supersymmetric contribution to  CP violation in
$B-\bar B$ mixing can be as large as the weak interaction
contribution\cite{CKLN}. Also new contributions to CP violating decay
amplitudes may arise with significant departures from the SM predictions.
As for FCNC phenomena in B physics, the model provides sizeable new
contributions to the mixing and the B radiative decays, but always keeping
below the experimental results.

Another possible constraint on new sources of CP violation comes from
Electric Dipole Moment (EDM) bounds on the neutron and on atoms.  Our 
model contains a massless up quark and thus there is no strong CP violation.
Though there are several new sources of CP violation, supersymmetric 
contributions to EDM's are sufficiently suppressed due to the large mass 
of the first two superpartner generations.

\section{Some Cosmological Considerations}

Supersymmetric models, where the messenger sector is identified with the
Froggatt-Nielsen sector and a single $U(1)$ symmetry is used both to give
large masses to the first two generations of sfermions and to generate the
flavor spectrum, are of considerable interest \cite{darkmatter} from the
cosmological point of view. Indeed, this class of low energy supersymmetry
breaking models naturally predicts (superconducting) cosmic strings
\cite{Riotto97}.  The presence of an Fayet-Iliopoulos  $D$-term $\xi^2$ induces
the spontaneous breakdown of the    $U(1)$ gauge symmetry  along some field
direction in the messenger sector.  Let us denote this field direction  
generically by $\varphi$ and its $U(1)$ charge by $q_\varphi$. In this case
local cosmic strings are formed whose mass per unit length is given by
$\mu\sim\xi^2$ \cite{Riotto97}. Since $\xi$ is a few orders of magnitude
larger than the weak scale, cosmic strings are not very heavy.  The crucial
point is that   some quark and/or the lepton superfields are charged under the
 $U(1)$ group.  Let us focus on one of the sfermion fields, $\widetilde{f}$
with generic $U(1)$-charge $q_f$, such that  sign $q_f=$ sign $q_\varphi$. The
potential for the fields $\varphi$ and $\widetilde{f}$ is written as
\begin{equation}
V(\widetilde{f},\varphi)=q_\varphi^2\mt |\varphi|^2 +
	q_f^2\mt|\widetilde{f}|^2 + 
	\frac{g^2}{2}\left(q_\varphi |\varphi|^2 + 
	q_f|\widetilde{f}|^2+\xi^2 \right)^2 +
	\lambda |\widetilde{f}|^4, 
\end{equation}

where we have assumed, for simplicity, that $\widetilde{f}$ is $F$-flat. The
parameter $\lambda$  is generated from the standard model gauge group $D$-terms
and  vanishes if we take $\widetilde{f}$ to denote a family of fields
parameterizing a $D$-flat direction.

At the global minimum $\langle \widetilde{f}\rangle=0$ and the electric charge,
the baryon and/or the lepton numbers are conserved. The soft breaking mass term
for the sfermion reads
\begin{equation}
\Delta m_{\widetilde{f}}^2=q_f\left(q_f - q_\varphi\right)\mt,
\end{equation}
and is positive by virtue of the hierarchy $q_\varphi<q_f<0$ (recalling
$\mt<0$). Consistency with experimental bound requires 
$\Delta m^2_{\widetilde{f}}$ to be of the
order of $\left(20\:{\rm  TeV}\right)^2$ or so,  which in turn requires
$\xi^2 \sim (4\pi/g^2)\widetilde{m}^2\sim (10^2\:{\rm TeV})^2$. Notice that
$\Delta m_{\widetilde{f}}^2$ does not depend upon $\xi^2$.

Let us analyze what happens in the core of the string. In this region of space,
the vacuum expectation value of the field vanishes,
  $\langle|\varphi|\rangle=0$,  and nonzero values of $\langle
|\widetilde{f}|\rangle$ are energetically preferred in the string core
\begin{equation}
\langle |\widetilde{f}|^2\rangle=
	\frac{- \mt q^2_f - g^2\xi^2 q_f}{g^2 q^2_f + 2\lambda}.
\end{equation}
Since the vortex is cylindrically symmetric around the $z$-axis, the condensate
will be of the form $\widetilde{f}=\widetilde{f}_0(r,\theta)\:{\rm
e}^{i\eta_f(z,t)}$, where $r$ and $\theta$ are the polar coordinate in the
$(x,y)$-plane. One can check easily that the kinetic term for $\widetilde{f}$
also allows a nonzero value of $\widetilde{f}$ in the string and therefore one
expects the existence of bosonic charge carriers inside the strings. The latter
are, therefore, superconducting.

These superconducting cosmic strings formed at temperatures within a few orders
of magnitude of the weak scale may generate primordial magnetic fields
\cite{Riotto97} and even give rise to the observed baryon asymmetry
\cite{br98}.  Indeed, during their evolution, the superconducting cosmic
strings  carry some baryon charge. The latter is efficiently  preserved from
the sphaleron erasure and may be released in the thermal bath  at low
temperatures.  In such a case, the charge carriers inside the strings are
provided by the scalar superpartner of the fermions that carry baryon (lepton)
number. Since these scalar condensates are charged under $SU(2)_L$, baryon
number violating processes are frozen in the core of the strings and the baryon
charge number can not be wiped out at temperatures larger than $T_{EW}\sim 100$
GeV. In other words, the superconducting strings act like ``bags'' containing
the baryon charge and protect it from sphaleron wash-out throughout the
evolution of the Universe, until baryon number violating processes become
harmless.  This mechanism is efficient even if the electroweak phase transition
in the MSSM is of the second order and therefore does not impose any upper
bound on the mass of the Higgs boson \cite{br98}.

\section{Conclusion}

We have presented a renormalizable model of low energy flavor and
supersymmetry breaking in which all mass scales are produced dynamically.
A U(1) gauge group mediates large contributions to the masses of the first two
generations of scalars, of order 20 TeV,  while suppressing the masses
of their fermionic partners.  Excessive FCNC are successfully avoided,
in part, by decoupling the scalars of the first two families.  CP
violation in the kaon system is also predicted to be within experimental
bounds.  We are able to produce the observed fermion masses and mixing angles
while maintaining perturbative unification of gauge coupling constants at 
${\rm M}_{\mathrm{GUT}}$.  However, we did not explicitly construct a 
complete model of flavon interactions having the correct vacuum, though 
we have made it plausible that one could be produced.

Our goal was to produce a model in which the sectors responsible for
scalar masses and fermion masses could be identified.  The resulting
model, as an unintended consequence, potentially  solves at least two major
problems of fundamental physics.  First, the model predicts a massless up 
quark.  This is the simplest viable solution to the strong CP problem.  
Second, the model predicts the existence of light superconducting cosmic 
strings, which could be the source of the magnetic fields that are observed 
on the cosmological scale.  These strings may also be responsible for the 
baryon asymmetry of the universe.

Our model suffers from the same '$\mu$-term' problem that exists in
most gauge mediated models \cite{dvalimu}.  We can naturally generate
a $\mu$-term via loop corrections if we include, for example, the
terms ${H}_u {\bf 10}^V {\bf 10}^V$, ${H}_d \bar{\bf 10}^V
\bar{\bf 10}^V$ and $\chi {\bf 10}^V \bar{\bf 10}^V$.  At one loop, a
$\mu$-term appears with coupling constant
$\mu\sim\frac{1}{16\pi^2}\frac{F}{M}$, where $\langle\chi\rangle = {M}
+ \theta\theta{F}$.  As pointed out by Dvali, Pomarol and Giudice
\cite{dvalimu}, the scalar coupling, $B_{\mu} {H}_u {H}_d$,
also appears at one loop with
$B_{\mu}\sim\frac{1}{16\pi^2}\frac{F^2}{M^2}\sim(4\pi\mu)^2$, which is
too large for natural electroweak symmetry breaking.  It may be
possible to adopt the mechanisms of reference \cite{dvalimu} to
suppress this $B_{\mu}$ term, or to produce acceptable $\mu$ and
$B_{\mu}$ terms via the mechanism of ref.~\cite{nilles}.

As the first  renormalizable and explicit example of the
Effective Supersymmetry \cite{CKN} approach to flavor and
supersymmetry breaking, this model reproduces the success of the
standard model in explaining the observed size of FCNC and absence of
Lepton Flavor Violation (LFV).
In fact this model is surprisingly successful, as the supersymmetric
contributions to CP violating effects in $K-\bar K$ mixing, which
even with 20 TeV squarks are potentially 100 times too large,  are
sufficiently small.
The CP violating phases  in $B_d-\bar B_d$  and $B_s-\bar B_s$ mixing
receive a large nonstandard contribution from left-handed bottom squark
exchange. It remains to be calculated whether any other nonstandard FCNC,
CP violating, and LFV effects are large enough to be revealed by new,
more stringent experiments.
\vskip 1.2cm

This work was supported in part by the Department of Energy Grant
No. DE-FG03-96ER40956 and by the European Network "BSM" 
No. FMRX CT96 0090 - CDP516055.  FL was supported in part by the U.S.
Department of Energy under Grant No. DOE-ER-40561. DEK would like to thank
D. Wright for useful discussions and G. Kane for helpful 
comments.  FL would like to thank D.B. Kaplan for useful discussions and 
helpful suggestions.  The authors would like to thank N. Nigro for his
copy editing expertise.

\appendix
\section{Flavor Dependent Contributions to Uncharged Squarks}
In this appendix, we show explicitly that leaving all three generations of
right-handed down squarks uncharged will not produce a degenerate spectrum.  To
see this, let us look at the matrix explicitly.  Filling in the remaining 
entries and rotating fields to simplify the matrix we have:
\begin{equation}
M^d=\bordermatrix{
        &\bar{d}_0^1&\bar{d}_0^2&\bar{d}_0^3
                &\bar{d}_{-b}^V&\bar{d}_0^{\bar V}\cr
        d_a^1 &0&0&0&0&\langle \phi_{-a} \rangle\cr
        d_b^2 &0&0&0&\hdv&\langle \phi_{-b} \rangle\cr
        d_0^3 &0&0&\hdv&0&0\cr
        d_0^V &0&\hdv&\hdv&0&\langle \chi_{0} \rangle\cr
        d_m^{\bar V} &\langle \phi_{-m} \rangle
                &\langle \phi_{-m}\rangle&\langle \phi_{-m} \rangle
                &\langle \chi_{b-m} \rangle&0\cr}
\end{equation}
(the effects of a non-zero (5,5) element are minimal, so it is set to zero).
 This matrix produces the following down-type Yukawa couplings in the low
energy theory:
\begin{eqnarray}
  \bar{d}^{g}
  \hdv
    \left( \begin{array}{ccc} 0&\sim\!\epsilon_{1}&\sim\!\epsilon_{1}\\
\sim\!\epsilon_{3}&\sim\!\epsilon_{2}+\sim\!\epsilon_{3}&\sim\!\epsilon_{2}+\sim\!\epsilon_{3}\\
      0&0&\sim 1\\
      \end{array} \right)_{gh}
  d^h,
\end{eqnarray}
where
\begin{displaymath}
\begin{array}{l}
\epsilon_{3}=\displaystyle\frac{\langle\phi_{-m}\rangle}{\langle\chi_{b-m}\rangle}.\\
\end{array}
\end{displaymath}
To see that, for example, the RH down squarks are not degenerate, we
examine the squark mass (squared) matrix.  For our purposes, we can ignore
terms proportional to $\hdv$ (which would be of the same order as the bottom
quark mass).  In this approximation, the relevant superpotential couplings are
\begin{equation}
W\supset \sum_{i=1}^3 \lambda_i {\bf\bar{5}}_0^i {\bf 5}_m^{\bar{V}} \phi_{-m}
+ \lambda_V {\bf\bar{5}}_{-b}^V {\bf 5}_m^{\bar{V}} \chi_{b-m} +
\lambda_{\bar{V}} {\bf\bar{10}}_e^{\bar{V}} {\bf 10}_0^V \chi_{0},
\end{equation}
and the mass squared matrix is:
\begin{eqnarray}
\left(\begin{array}{ccccc} \left|\lambda_1 \right|^2 \left|
\phi\right|^2&\lambda_1^* \lambda_2 \left| \phi\right|^2&\lambda_1^* \lambda_3
\left| \phi\right|^2&\lambda_1^* \lambda_V \phi^* \chi&0\\
\lambda_2^* \lambda_1 \left| \phi\right|^2&\left|\lambda_2 \right|^2\ \left|
\phi\right|^2&\lambda_2^* \lambda_3 \left| \phi\right|^2&\lambda_2^* \lambda_V
\phi^* \chi&0\\
\lambda_3^* \lambda_1 \left| \phi\right|^2&\lambda_3^* \lambda_2 \left|
\phi\right|^2&\left|\lambda_3 \right|^2 \left| \phi\right|^2&\lambda_3^*
\lambda_V \phi^* \chi&0\\
\lambda_V^* \lambda_1 \chi^* \phi&\lambda_V^* \lambda_2 \chi^* \phi&\lambda_V^*
\lambda_3 \chi^* \phi&\left|\lambda_V \right|^2 \left| \chi \right|^2&0\\
0&0&0&0&\left|\lambda_{\bar{V}} \right|^2 \left| \chi_{0} \right|^2\\
\end{array}\right)
\end{eqnarray}
where, for simplicity, $\phi = \phi_{-m}$ and $\chi = \chi_{b-m}$.  This matrix
has three zero eigenvalues and two eigenvalues of order
${\langle\chi\rangle}^2$.  The three generations of squarks receive degenerate
weak-scale contributions to their masses from the two loop diagrams in Figure 1
of \cite{DNS}.  However, the FN fields receive large \susy breaking
contributions to their masses (of order $\sqrt\mt$).  When $\mt$ is added to
the (4,4) component of the matrix, there is one less zero eigenvalue.  For
$\mt\lsim\xi^2 \sim {\langle\chi\rangle}^2$ and ${\langle\phi\rangle}\ll
{\langle\chi\rangle}$, this matrix has two eigenvalues of order
${\langle\chi\rangle}^2$, and one of order
$\left(\frac{\langle\phi\rangle}{\langle\chi\rangle}\right)^2 \mt =
\epsilon_{3}^2 \mt$.  In order to produce the correct mass ratios and mixing
angles without significant fine tuning, it turns out that $\epsilon_3$ must be
of order $10^{-2}$.  For $\mt\sim (20\TeV)^2$, the third eigenvalue is of order
$(200\GeV)^2$, thereby destroying the weak-scale degeneracy.  A more careful
analysis reveals additional flavor-dependent contributions at one loop.  In
fact, the only way to protect this degeneracy is to require all of the SU(5)
multiplets (and hence, all flavons) to be uncharged under \u1x, clearly a
useless choice.

\end{document}